\documentclass{article}

\usepackage[utf8]{inputenc}
\usepackage[T1]{fontenc}
\usepackage{amsmath, amssymb} 
\usepackage{graphicx} 
\usepackage[table]{xcolor} 
\usepackage[colorlinks=true, linkcolor=black, citecolor=black, urlcolor=black]{hyperref}
\usepackage{natbib} 
\usepackage{caption}
\usepackage{moreverb,url}
\usepackage{tikz} 
\usepackage{soul} 

\usepackage{rotating} 
\usepackage{array} 
\usepackage{booktabs} 

\newcommand\BibTeX{{\rmfamily B\kern-.05em \textsc{i\kern-.025em b}\kern-.08em
T\kern-.1667em\lower.7ex\hbox{E}\kern-.125emX}}

\title{Missing Data Imputation in the Context of Propensity Score Analysis: A Systematic Review}

\author{%
Saghar Garayemi\thanks{Department of Mathematics, Institute of Mathematical Statistics and Artificial Intelligence in Medicine, University of Augsburg. Corresponding Author: \texttt{sagha.garayemi@uni-a.de}} 
\and
Reza Ali Akbari Khoei\thanks{Road Traffic Injury Research Center, Tabriz University of Medical Sciences, Tabriz, Iran}
\and
Sarah Friedrich\footnotemark[1]%
}
\date{} 

\begin{document}

\maketitle

\begin{abstract}
Missing data is a common challenge in observational studies. Another challenge stems from the observational nature of the study itself: Here,
propensity score analysis can be used as a technique to replicate conditions similar to those found in clinical trials. With regard to the missing data, a majority of studies {only analyze the complete cases, but this has several pitfalls.}
In this review, we investigate which methods are used for the handling of missing data in the context of propensity score analyses.
Therefore, we searched PubMed for the keywords `propensity score' and `missing data', restricting our search to the time between January 2010 and February 2024. The PRISMA statement was followed in this review. A total of 147 articles were included in the analyses.

A major finding of this study is that although the usage of multiple imputation (MI) has risen over time, only a limited number of studies describe the mechanism of missing data and the details of the MI algorithm.

\vspace{1em}
\end{abstract}

\section{Introduction} \label{sec:intro}
\noindent When looking for a causal relationship, the ideal design is a randomized controlled trial (RCT)\cite{austin2011introduction}. In practice, however, it is not always feasible to conduct an RCT. Two important reasons are the willingness of people to participate (ethical issues) and feasibility issues, such as costs. Observational studies constitute an alternative in situations where an RCT is not possible\cite{west2008alternatives}.
These data often come from active registries and usually have a rather large number of observations, which are ideally gathered at a predetermined time and according to some protocol. However, since this data is usually not collected for the specific research question at hand, quality is often lower compared to data extracted from an RCT \cite{valojerdi2018brief}. This poses two challenges: First, special methods need to be employed in order to derive causal conclusions. And second, missing data is a common problem due to unstructured data collection.

For the purpose of this review, we restrict ourselves to articles using propensity score methods to deal with the observational nature of the data.

The propensity score (PS) is the probability of treatment assignment for each individual given their respective covariates\cite{rosenbaum1983central,Rosenbaum1984}.
The most common methods for propensity score adjustment include matching, PS covariate adjustment, inverse probability of treatment weighting (IPTW) and stratification\cite{austin2011introduction,friedrich2020causal}. Commonly, propensity scores are estimated using logistic regression, but other approaches have been suggested as well\cite{gharibzadeh2018comparing}.

Moreover, the handling of missing data might also affect the estimation of propensity scores \cite{choi2019comparison}.
Three different types of missing data are distinguished in the literature, affecting the method that can be used for analyzing the data\cite{gelman2007data}:

\begin{itemize}
\item \emph{Missing Completely at Random (MCAR):}
Missing data are called missing completely at random (MCAR) if the individuals with the missing data are a random subset of the entire sample of subjects.
\item \emph{Missing Not At Random (MNAR):}
Missing data is called missing not at random (MNAR) when the probability that an observation is missing is influenced by information not observed, such as the observations (unobserved) value.
\item \emph{Missing At Random (MAR):}
When missing data is neither MCAR nor MNAR, it is called Missing At Random (MAR). In this case, the probability that an observation is missing often depends on the available information for that subject, meaning that the reason for the missingness is related to other observed patient characteristics. In this context, the missing data can be considered random, conditional on the observed patient characteristics\cite{little2019statistical}.
\end{itemize}

Different methods exist for dealing with missing data. In general, it is not recommended to exclude the subject with missing data and use only complete cases for analysis, since this might introduce bias\cite{haneuse2016learning}. Instead, missing data should be imputed to avoid reducing the sample size and data quality. In general, imputing missing data implies filling in the missing data with a suitable replacement such that the model is still valid\cite{schafer1997analysis}. Many methods to impute missing data exist in the literature, for example simple imputation, mean imputation\cite{little2019statistical}, multiple imputation\cite{little2019statistical}, K-NN imputation\cite{batista2003analysis} and random forest imputation\cite{stekhoven2012missforest}.

The aim of this systematic review is to assess how missing data is handled in the context of propensity score analyses. In particular, we focus on which methods of imputation are being used and how well papers stick to the STROBE\footnote{Strengthening the Reporting of Observational Studies in Epidemiology} guideline for reporting missing data.

\section{Methods}

In this systematic review, clinical studies with missing data using some form of propensity score methods were included. To specifically focus on the applied context, we excluded animal studies and papers with a purely methodological focus. The systematic review followed the Preferred Reporting Items for Systematic Reviews and Meta-Analyses (PRISMA) statement \cite{moher2010preferred}.One researcher (SG) conducted the first screening of publications, reviewing the list of all retrieved papers to evaluate eligibility based on the inclusion and exclusion criteria. In some cases, evaluating the title alone was sufficient, while in others, the abstract was also considered. When necessary, the entire text of the article was evaluated. In times of ambiguity, two other researchers were brought into the decision-making process to reach an agreement.

\subsection{Systematic literature search and study selection}
\begin{figure}[htp]
\centering
\begin{tikzpicture}[>=latex, font={\sf \scriptsize}]
\tikzstyle{bluerect} = [rectangle, rounded corners, minimum width=0.8cm, minimum height=0.8cm, text centered, draw=black, fill=cyan!60!gray!45!white, rotate=90, font=\sffamily]
\tikzstyle{roundedrect} = [rectangle, rounded corners, minimum width=7cm, minimum height=0.8cm, text centered, draw=black, font=\sffamily]
\tikzstyle{textrect} = [rectangle, minimum width=3cm, text width=3.6cm, minimum height=0.8cm, draw=black, font={\sffamily \scriptsize}]
\node (top1) at (-1, 10.7cm) [draw, roundedrect, fill=yellow!80!red!70]
 {\textbf{Identification of studies via database and registers}};
\node (r1blue) at (-5.5cm, 8.3cm) [draw, bluerect, minimum width=2cm]{Identification};
\node (r1left) at (-2.8cm, 8.5cm) [draw, textrect, minimum height=2cm]
{\small Records identified from:

Databases ($n=225$) \\
 };
\node (r1right) at (1.5cm, 8.5cm) [draw, textrect, minimum height=3cm]
 {\small Records removed before screening: \\

Systematic Reviews/Reviews/Literature Reviews ($n=5$) \\
Protocols ($n=3$) \\
Non-English($n=1$)\\Repeated ($n=2$)

 };
\node (r2blue) at (-5.5cm, 4cm) [draw, bluerect, minimum width=4.75cm]
 {Screening};
\node (r2left) at (-2.8cm, 5.8cm) [draw, textrect, minimum height=1cm]
 {\small Records screened ($n=216$)};
\node (r2right) at (1.5cm, 5.8cm) [draw, textrect, minimum height=1cm]
 {\small Number of articles excluded with reason: \\
 Theoretical papers ($n=59$) \\   Animal Study ($n=1$)};
\node (r3left) at (-2.8cm, 3.6cm) [draw, textrect, minimum height=1cm]
 {\small Number of articles assessed for eligibility ($n=156$)};
\node (r3right) at (1.5cm, 3.6cm) [draw, textrect, minimum height=1.5cm]
{\small Records excluded: \\

No access to the full paper ($n=1$)\\
 Preprint ($n=1$)};
\node (r4left) at (-2.8cm, 1.3cm) [draw, textrect, minimum height=1cm]
 {\small Reports assessed for eligibility ($n=154$)};
 \node (r4right) at (1.5cm, 1.3cm) [draw, textrect, minimum height=1cm]
 {\small Number of full-text articles excluded with a reason: \\ No Medical Context ($n=4$) \\Practical Guide ($n=1$)\\};
\node (r5blue) at (-5.5cm, -0.5cm) [draw, bluerect, minimum width=1.5cm]
 {Included};
\node (r5left) at (-2.8cm, -0.5cm) [draw, textrect, minimum height=1cm]
{\small Studies included in review
$n=147$ \\
};
\draw[thick, ->] (r1left.east) -- ++(0.2cm,0) -- (r1right.west);
\draw[thick, ->] (r1left) -- (r2left);
\draw[thick, ->] (r2left.east) -- ++(0.2cm,0) -- (r2right.west);
\draw[thick, ->] (r2left) -- (r3left);
\draw[thick, ->] (r3left.east) -- ++(0.2cm,0) -- (r3right.west);
\draw[thick, ->] (r3left) -- (r4left);
\draw[thick, ->] (r4left.east) -- ++(0.2cm,0) -- (r4right.west);
\draw[thick, ->] (r4left) -- (r5left);
\end{tikzpicture}
\caption{ Flowchart of the Study selection Process.}
 \label{fig:prisma}
\end{figure}
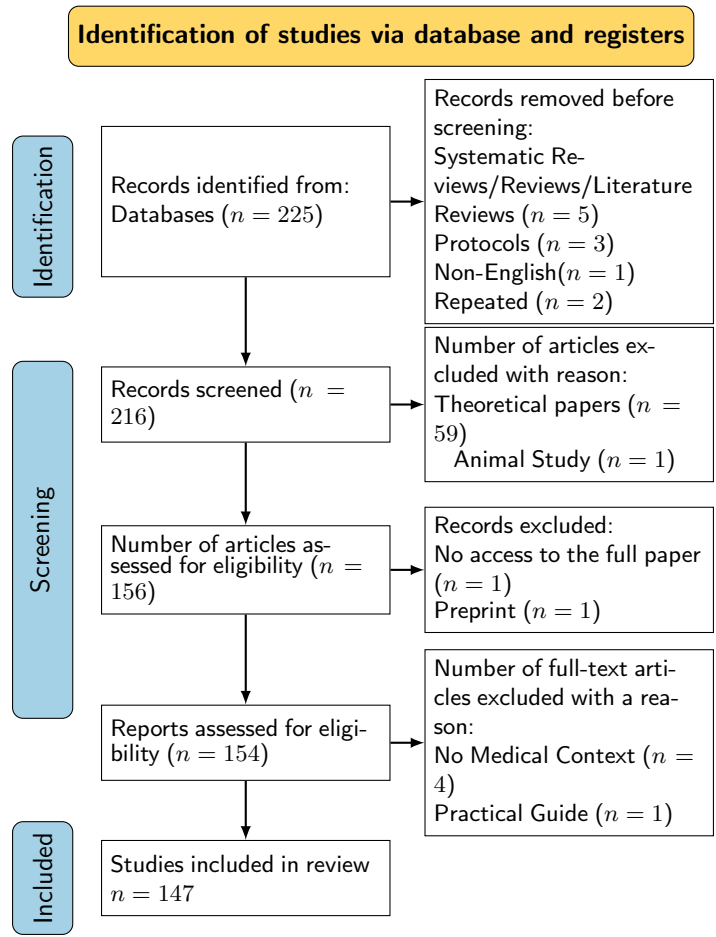

We performed a literature search using PubMed with the search terms `(Propensity Score) AND (Missing Data)' and restricted the search to publications from January 2010 to January 2024.
This search resulted in 225 manuscripts. The date of the last search was February 5th, 2024.
We excluded articles not in English as well as systematic reviews and purely theoretical papers. The search was restricted to human studies, and one article had to be excluded since no access to the full text was possible. The PRISMA flow chart is depicted in Figure~\ref{fig:prisma}.

\subsection{Data Extraction}
We extracted data on study characteristics such as study design, year of publication, number of treatment groups and location. Moreover, we considered variables related to the missing data,
such as imputation methods used, pre- and post-imputation outcome comparisons, and the use of sensitivity analyses. We also considered items that are important in the STROBE guideline such as the proportion of missing data and the reason for missingness. Data extraction used the same procedure as title screening, with one researcher performing the initial extraction and two more researchers involved in cases of uncertainty to establish an agreement. Reporting the proportion of missing data is a fundamental but critical component of research transparency. It sheds light on the quantity of missing data and aids in determining the potential bias it may introduce. The reason why missing data happens is critical for determining potential bias and guaranteeing the quality of study findings. The reason for missing data can influence the statistical methods used, as well as the conclusions' internal validity and generalizability. For example, if data is consistently absent (e.g., sicker patients skip out), the results may be skewed. Addressing the causes of missing data helps ensure that proper methods are utilized to handle it and that the study's conclusions are robust and applicable to larger populations.
If (multiple) imputation was used, we furthermore extracted variables related to the imputation procedure such as the number of imputations and the variables used in the MI model, as this
influences the stability of the imputed data and aids reproducibility.
We also considered if sensitivity analyses are undertaken to ensure the results are robust to various assumptions about missing data. This is a vital stage in guaranteeing the validity of the study conclusions.
Finally, we considered the method used to estimate the propensity score model after dealing with the missing data.

\subsection{Scientometric Analysis}
We conducted a scientometric analysis to investigate how the context, we focused on, is regarded in research over time.  Utilizing the PubMed Library and the final list of papers retrieved between 2010 and 2023, we used the bibliometrix package in \texttt{R}\cite{r2013r} to do a scientometric analysis. The program uses each paper's citation to create the final plot\cite{aria2017bibliometrix}.

\section{Results}
\subsection{Study characteristics}

\begin{figure}[htp]
\includegraphics[width=1\linewidth]{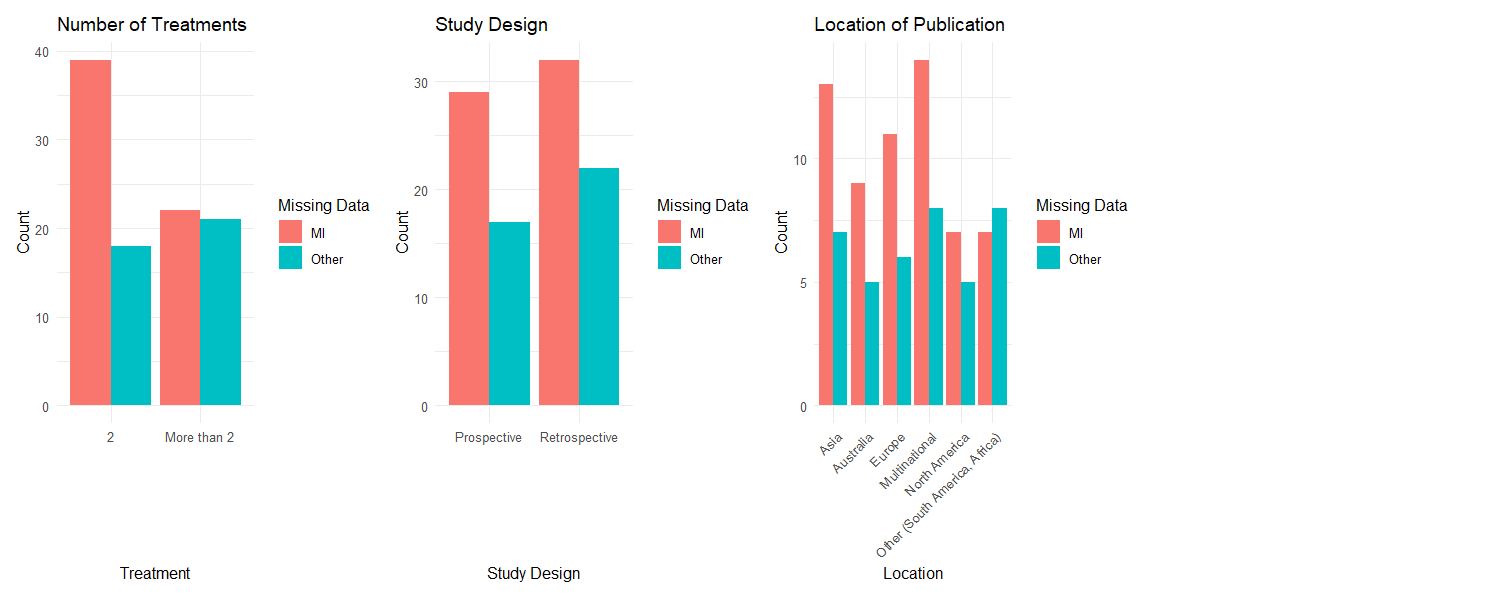}
\caption{Study Characteristics.}
 \label{Characteristics}
\end{figure}

Figure~\ref{fig:prisma} depicts the PRISMA flowchart.
In the first step, we identified 225 papers which were screened for eligibility. After all exclusion and inclusion criteria were applied, 147 papers remained for the final analyses. 
Table~\ref{Table 1} provides an overview of the extracted variables, while the study characteristics are summarized in Table~\ref{Table 2}. From Table~\ref{Table 2} we see that 104 (70.74\%) of the papers collected the data retrospectively from which 57 (54.80\%) used MI. On the other hand, 43 (29.25\%) prospectively gathered the data; from them, 26 (60.46\%) used MI methods to handle missing data. 135 (91.83\%) articles compared two treatment groups while only 12 (8.16\%) used data from more than two treatment groups, see also Figure~\ref{Characteristics}.

To identify possible time trends, publications were divided into 3-year intervals 2010-2012, 2013-2015, 2016-2018, 2019-2022, and 2023-2024. Note that the last interval contains only one complete year, which should be kept in mind when interpreting the results. 
 From Table~\ref{Table 4} we can see that most papers were published between 2022 and 2024, and there is an increasing trend regarding the number of published papers.
 
 Considering the sensitivity analysis, we noticed that less than half of the papers using MI performed a sensitivity analysis. Most of the papers (82\%) used logistic regression to estimate the propensity scores. Figure~\ref{doublepie} briefly shows the distribution of all methods used for estimating the propensity scores.
Regarding the location where the data was gathered, Table~\ref{Table 2} shows that 60 (40.81\%) papers were based on data from North America (US and Canada). After that, Europe and Asia had almost the same percentage (38 (25.85\%) and 36 (24.48\%) papers, respectively).
There were also some papers from Australia, South America, and Africa.
73.68\% of papers originating in European countries used MI for imputing missing data and 32 (53.33\%) of the papers belonging to North America used MI for this purpose.

\begin{table}[htp]
\scriptsize\sf\centering
\caption{List of Variables Extracted.\label{Table 1}}
\begin{tabular}{@{} p{6cm} p{2.5cm} @{}}
\toprule
\rowcolor{lightgray}
Variable & Response Option \\
\midrule
Article Characteristics & \\
\midrule
Year Of publication  & - \\
Title of publication & - \\
Number of treatment groups compared &2/\(>2\) \\
\midrule
Missing data method & \\
\midrule
Proportion of missing data reported  & Yes/No \\
Missing data imputation method reported & Yes/No \\
Missing data mechanism (MAR, MCAR) mentioned & Yes/No \\
 Reason for missing data given  & Yes/No \\
 Missing data sensitivity conducted  & Yes/No \\
 Analysis compared between those with complete and incomplete data  & Yes/No \\
 Variables included in MI explained (if MI used) & Yes/No \\
 Number of imputations specified (if MI used)  & Yes/No \\
 Methods used to estimate propensity scores after MI & - \\
 Location of Publication & - \\
\bottomrule
\end{tabular}
\end{table}

\subsection{Detailed Report of Missing Data}
In total, 136 (93.79\%) papers mentioned the amount of missing data. Missing data mechanisms, i.e.~whether data was assumed to be MAR, MCAR or MNAR were only reported in 36 (24.82\%) papers. Also, only 45 (31.03\%) of the papers stated a reason for why data was missing.

\subsection{Missing Data: Sensitivity Analysis}
We only considered sensitivity analyses, where the result of the complete case analysis was compared to the analysis based on imputed data. Of the papers that used MI, more than half 56 (67.47\%) performed a sensitivity analysis. However, none of them considered the possibility of MCAR and only 11
(13.25\%) mentioned the MAR assumption. In total, 18 (21.68\%) articles reported different results after the sensitivity analysis.

\subsection{Missing Data Methods}
\begin{figure}[htp]
 centering
 \includegraphics[width=1\linewidth]{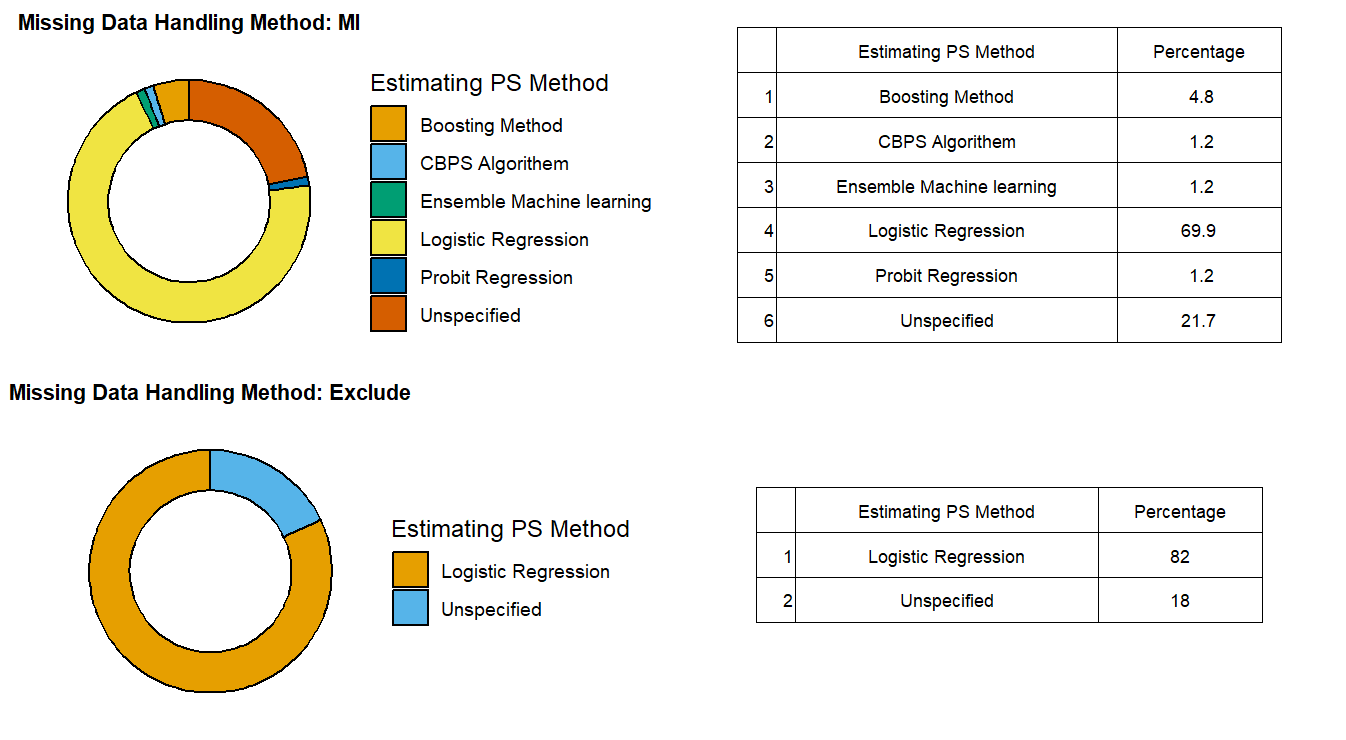}
 \caption{Distribution of Propensity Score Estimation Methods}
 \label{doublepie}
\end{figure}

\begin{figure}[htp]
\centering
\includegraphics[width=0.8\textwidth]{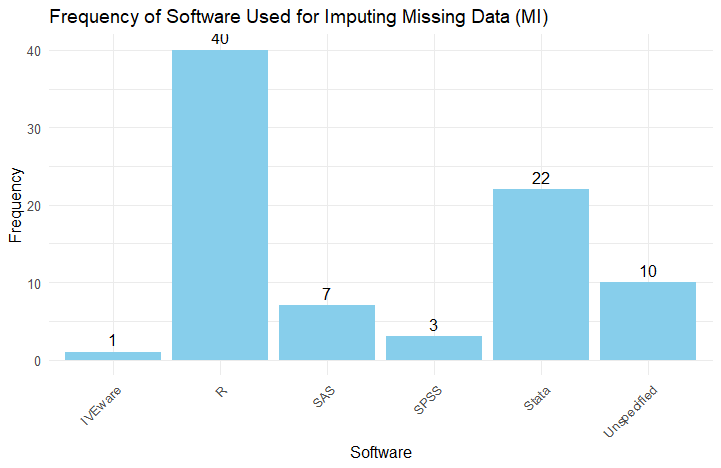}
 \caption{Software Used for Imputing Missing Data}
\label{software}
 \hfill
\end{figure}

83 (57.24\%) papers utilized MI for imputing missing data, while 50 (34.48\%) restricted their analyses to complete cases. The rest of the papers used other methods like interpolation ($n=1$) or mean imputation ($n=1$). Some also used random forest ($n=2$) or regression imputation ($n=1$). Table~\ref{table3} shows the complete list of other methods used with corresponding frequencies. When missing data was imputed, this was done using a variety of software. As depicted in Figure~\ref{software}, \texttt{R} was the most common software for this purpose and used in 40 (48.19\%) articles. Stata ($n=22$, 26.50
As shown in Figure~\ref{doublepie} logistic regression is the most common method to estimate propensity scores, both for the group that used MI for imputing missing data and the one excluding missing data. Interestingly, when MI was used, sometimes other approaches to propensity score estimation were reported as well, for example boosting-based methods. One paper also used the CBSP (Covariate Balance Propensity Scores) Algorithm\cite{imai2014covariate} (using a package in \texttt{R} having the same name). This was not the case in papers using complete case analyses only. 

\begin{table}[htp]
\scriptsize\sf\centering
\caption{Study Characteristics. Results are given as absolute numbers and $n(\%)$.\label{Table 2}}
\begin{tabular}{@{}p{4cm}p{1.5cm}p{1.5cm}p{1.5cm}@{}}
\toprule
\rowcolor{lightgray}
Study Characteristics & MI used & MI not used & Total\\
\midrule
  Design & \\
 Retrospective &  57 & 47 & 104 (70.74\%)  \\
 prospective &  26 & 17 & 43 (29.25\%) \\
\midrule
 Number of treatment groups compared& \\
 2&  74 & 61 & 135 (91.83\%)\\
\(>2\)&  9 & 3 & 12 (8.16\%)\\
\midrule
 Location of publication& \\
 Asia& 15 & 21 & 36 (24.48 \%)\\
 Europe&    28 & 10 & 38 (25.85 \%)\\
 North America&   32 & 28 & 60 (40.81 \%)\\
 Australia&  3 & 1 & 4 (2.72 \%)\\
 multinational & 3 & 2 & 5 (3.4 \%)\\
 Other(South America, Africa)&        2 & 2 & 4 (2.72\%)\\
\bottomrule
\end{tabular}
\end{table}

\subsection{Following STROBE Guidelines}
One of the study's purposes was to determine how well publications adhere to the STROBE guidelines.
We found that 45 (31.03\%) publications explained the cause for missing data. The proportion of missing data was indicated by 136 (93.79\%) papers, and the approach for handling missing data was mentioned in all but one study.  Overall, only 17 (11.56\%) articles examined all item issues in the STROBE standards.

\subsection{Scientometric Analysis Results}
\begin{figure}[htp]
\centering
\includegraphics[width=0.85\linewidth]{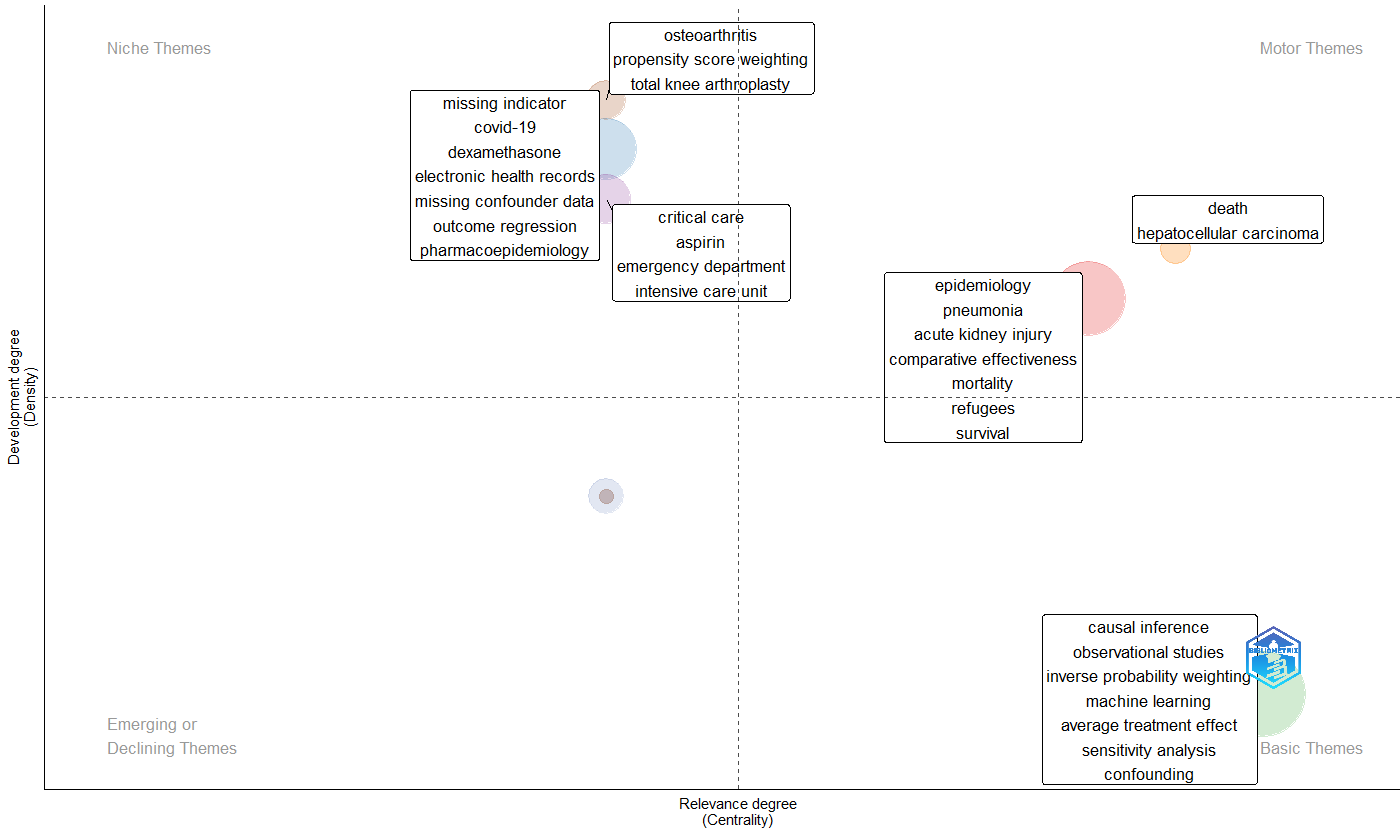}
\caption{Thematic Presentation of Emerging Trends and Knowledge}
 \label{sciento}
\end{figure}

One goal of doing a systematic review is to identify gaps in the study area while simultaneously generating fresh ideas for future research. A scientometric study is useful for identifying areas that require additional attention.
Figure~\ref{sciento} depicts the results based on our literature search. The plot consists of four parts: 
`Motor themes' show well-developed primary topics, which still leave room for future investigation. `Niche themes' represent themes that are well developed and isolated, i.e.~of limited importance to the research field\cite{cobo2011approach}.
`Basic themes' comprise foundational topics that still require additional investigation. Finally, Emerging or Declining Themes discuss subjects that are rather marginal and weakly developed. According to the plot, there are seven basic themes where there is insufficient study. This includes causal inference, observational studies, inverse probability weighting, machine learning, average treatment effect, sensitivity analysis, and confounding (the order of the themes is irrelevant).
We also discovered that sensitivity analysis is sometimes overlooked, thus it should be done more frequently when missing data is imputed. Causal inference and observation studies are the underlying concepts that drive the usage of propensity scores, especially in non-randomized studies when researchers seek to quantify causal effects. Identifying them as core topics highlights their importance in establishing a framework for the subject of our systematic study. The scientometric analysis found  `Epidemiology', `Pneumonia', `Acute Kidney Injury', and `Hepatocellular Carcinoma' as Motor themes in the overall research landscape. These issues are both central and well-developed, reflecting their importance in modern medical research. The use of propensity score approaches, particularly in the context of missing data, seems to be of great relevance in these fields. 

\begin{table}[htp]
\scriptsize\sf\centering
\caption{List of Other Methods for imputation\label{table3}}
\begin{tabular}{@{} p{5cm} p{0.2cm} @{}}
\toprule
\rowcolor{lightgray}
 Method  & Count \\
\midrule
 Random Forest&2\\
Add Category `Missing' to the Variable & 1 \\
IPW   & 1 \\
Mean Imputation  & 1 \\
Mean Imputation and Exclusion &1\\
Missing Interpolation& 1 \\
Multivariate Normal Regression &1\\
 Not clear &1\\
 PS Matching  &1\\
 Regression Imputation &1\\
\bottomrule
\end{tabular}
\end{table}

This is consistent with the focus of our systematic review, emphasizing the necessity of overcoming missing data difficulties to ensure valid and trustworthy results in research of these frequent disorders. The systematic review reveals the practical value of approaches, particularly missing data in propensity score analyses, in research areas like pneumonia, acute renal damage, and hepatocellular cancer. These findings underscore the need for ongoing methodological innovations for high-quality research.
The theme `Refugee' emphasizes research on refugee populations, focusing on health outcomes, access to care, and well-being.
Finally, `survival', `mortality' and `death' as Motor topics suggest that mortality research is critical to the field.

\captionsetup[figure]{skip=5pt}

\section{Discussion}
In this systematic review, we investigated which methods for handling missing data are employed in the context of propensity score analyses. In particular, we focused on whether the authors followed the STROBE guidelines, which methods are commonly used for imputing missing data and subsequently estimating the propensity score.

Although several guidelines encourage revealing the details of the imputation methods, our review found that out of 147 papers using propensity scores, 83 (57.24\%) employed MI for imputing missing data. However, only 17 (11.56\%) of them thoroughly detailed the method of data imputation, for example by providing the number of imputations, the variables employed, and the quantity of missing data.
 
Our review shows that the use of MI has increased over the last couple of years. This is congruent with the findings of Malla et al's systematic review\cite{malla2018handling}, which used MEDLINE and EMBASE databases as sources, and Hayati et al's systematic review\cite{hayati2015rise}, which used the Lancet and the New England Journal of Medicine as references. Nonetheless, Hayati's paper did not focus solely on observational studies.

An interesting finding is that among papers excluding missing data, only 24 (48\%) articles in our review acknowledged the possibility of bias in their findings. However, there seems to be a growing trend in at least reporting the result of sensitivity analyses as advised in the literature \cite{little2019statistical,sterne2009multiple}.

This systematic review focuses solely on medical research, and the conclusions should not be applied to other domains. Future research could go into bigger topics by including studies from other fields and evaluating more sources of research papers. Furthermore, doing a meta-analysis to analyze the impact of elements such as the missingness mechanism or the imputation method could be a worthwhile path for further inquiry.
Although we have seen that following STROBE rules is becoming more common, there are still some shortcomings in this regard. Researchers should include details about the settings they used to impute missing data so that it is easy to determine how trustworthy the results are.

The list of papers and the PRISMA checklist are in the supplementary material section.


\begin{sidewaystable}[ht]
\scriptsize\sf\centering
\caption{Time trend of variables related to Missing Data. Displayed are the number of papers $n(\%)$ for each category. Note that some percentages don't add up to 100 due to tiny categories, which were excluded for simplicity. 
}\label{Table 4}

\begin{tabular} {@{} p{6cm} p{3cm}p{1.5cm}p{1.2cm}p{1.2cm}p{1.2cm}|p{1.5cm} @{}}
 \hline
 \rowcolor{lightgray}
Variable & 2010-2012&2013-2015&2016-2018&2019-2021&2022-2024& Total$^*$\\
 \hline
 Publications (\(\%\)) &6(4.81)&19(12.92)&27(18.36)&46(31.29)&49(33.33)&147\\
  \hline
  \hline
 Multiple imputation&2(2.41) &14(16.87)&19(22.89)&23(27.71)&25(30.12)&83(57.24)\\
 \hline
 Exclude Missing Data&4(8)&3(6)&6(12)&19(36)&18(38)&50(34.48)\\
 \hline
 Other Method of Imputation&0&2(14.29)&2(14.29)&4(28.57)&6(42.86)&14(8.04)\\
 \hline
 Missing data mechanism mentioned &1(0.68)&6(4.13)&7(4.82)&16(11.03)&6(4.13)&36(24.82)\\
 \hline
 Reason for missing data given  &0(0)&11(7.58)&6(4.13)&20(13.79)&8(5.51)&45(31.03) \\
 \hline
Percentage of missing data given &5(3.44)&14(9.65)&22(15.12)&40(27.58)&19(13.10)&136(93.79)\\
\hline

 \textbf{Papers using MI} & & & & & & \textbf{83}\\
 \hline 
 Missing data sensitivity conducted &0(0)&9(11.11)&12(13.6)&26(19.8)&9(14.8)&56(67.47)\\
 \hline
Number of variables in the MI model mentioned &1(1.20)&10(12.04)&13(15.55)&20(24.09)&18(21.68)&62(74.69)\\
  \hline
Number of imputations specified &1(1.20)&10(12.04)&14(16.86)&19(22.89)&15(18.02)&59(71.08)\\

  \hline
  \multicolumn{7}{@{}p{17cm}@{}}{* Percentages in this column refer to the overall number of papers, i.e.~147 for all manuscripts and 83 for manuscripts using MI, respectively.} \\
\end{tabular}
\end{sidewaystable}

\bibliographystyle{plainnat} 
\bibliography{ref.bib} 

\end{document}